\documentstyle[prl,twocolumn,aps,epsf]{revtex}
\begin{document}

\hoffset-1cm

\draft
\preprint{TRP-96-??; NUCL-TH/960???}

\title{
A new class of Hanbury-Brown/Twiss parameters
}

\author{Urs Achim Wiedemann and Ulrich Heinz}

\address{
   Institut f\"ur Theoretische Physik, Universit\"at Regensburg,\\
   D-93040 Regensburg, Germany
}

\date{\today}

\maketitle

\begin{abstract}
In heavy ion collisions resonances can create strong non-Gaussian
effects in the 2-pion correlation data. Hence, the commonly used 
Gaussian fit parameters do not fully characterize these correlators.
We suggest a different set of HBT parameters which does not presuppose a 
particular shape of the correlator and allows to extract additional 
(non-Gaussian) information. Within a simple model for an expanding 
source including resonance decays it is shown that this additional 
information provides a clean distinction between scenarios with 
and without transverse flow.
\end{abstract} 



The aim of Hanbury-Brown/Twiss (HBT) interferometric analyses of heavy 
ion collisions is to
extract from the measured 2-particle correlators $C({\bf q},{\bf K})$
of identical particles as much information as possible about the
spatio-temporal distribution $S(x,K)$ of the particle emitting sources
in the collision region. It is based on the relation \cite{S73,CH94} 
 \begin{eqnarray}
   C({\bf p}_1,{\bf p}_2) &=&
   1 + {\left\vert \int d^4x\, S(x,K)\,
   e^{iq{\cdot}x}\right\vert^2 \over
   \left\vert \int d^4x\, S(x,K)\right\vert^2 } \, ,
   \nonumber \\
   q &=& p_1 - p_2\, ,
   \qquad K = \textstyle{1\over 2}(p_1 + p_2)\, .
 \label{1.1}
 \end{eqnarray}
So far, the interplay between the experimentally measurable 
momentum correlator $C({\bf q},{\bf K})$ and the theoretical 
concept of space-time emission function $S(x,K)$ was investigated 
mainly in the context of Gaussian approximations for both. 
Experimental data are commonly fit to a Gaussian ansatz~\cite{CSH95}
 \begin{mathletters}
  \label{1.2}
 \begin{eqnarray}
  \label{1.2a}
    C({\bf q},{\bf K}) &=&
    1 + \lambda({\bf K})\, e^{- q_i q_j R_{ij}^2({\bf K})}\, ,\\
    \label{1.2b}
        R_{ij}({\bf K}) &=& \left(
        \begin{array}{ccc}
                R_o^2    & R_{os}^2 & R_{ol}^2 \\
                R_{os}^2 & R_s^2    & R_{sl}^2 \\
                R_{ol}^2 & R_{sl}^2 & R_l^2
        \end{array}\right)\,, \qquad i,j = o,s,l\, .
  \end{eqnarray}
  \end{mathletters}
Here, the relative momentum components are defined parallel to the beam
($l$ = {\it longitudinal}), parallel to the transverse component of 
${\bf K}$ ($o$ = {\it out}), and in the remaining third direction
($s$ = {\it side}). $R_{os} = R_{sl} = 0$ for azimuthally symmetric 
collisions.

The fit parameters $R_{ij}^2({\bf K})$ can be calculated from a 
Gaussian ansatz for the emission function $S(x,K)$ in terms of 
space-time variances $\langle \tilde x_\mu \tilde x_\nu \rangle$
\cite{CNH95,WSH96}:
  \begin{mathletters}
    \label{1.3}
  \begin{eqnarray}
    \label{1.3a}
        S(x,K) &\simeq& N(K) \,S(\bar{x},K)\,
        e^{-{1\over 2} \tilde{x}^{\mu} \tilde{x}^{\nu}
        B_{\mu\nu}({\bf K})} \, ,\\
    \label{1.3b}
        (B^{-1})_{\mu\nu} &=&  \langle \tilde{x}_{\mu}
                               \tilde{x}_{\nu}\rangle, \  
        \tilde{x}_\mu = x_\mu - \bar x_\mu, \ 
        \bar x_\mu = \langle x_{\mu}\rangle,\\
    \label{1.3c}
        \langle f(x)\rangle &=& \langle f(x)\rangle (K) = 
        {\int d^4x\, f(x)\, S(x,K)\over\int d^4x \, S(x,K)} \, .
  \end{eqnarray}
  \end{mathletters}
In the resulting expression for the correlator, 
  \begin{equation}
     \label{1.4}
        C({\bf q},{\bf K}) = 1 + 
        \exp\left( - q^{\mu}q^{\nu} 
                   \langle \tilde x_\mu \tilde x_\nu \rangle \right) \, ,
  \end{equation}
only three of the four relative momentum components are independent, 
the fourth being fixed via the on-shell constraint $q^0 = {\bf q}
\cdot ({\bf K}/K^0) = {\bf q}\cdot\bbox{\beta}$. In the Gaussian 
framework, the consequences of the on-shell constraint
are well studied. Especially, the HBT radius parameters 
read~\cite{CSH95,HB95}
 \begin{equation}
   R_{ij}^2({\bf K}) = \langle (\tilde{x}_i-{\beta}_i\tilde{t})
                     (\tilde{x}_j-{\beta}_j\tilde{t})\rangle\, .
   \label{1.5}
 \end{equation}
This is the starting point of most spatio-temporal interpretations 
of correlation data. Note however that (\ref{1.6}) presupposes a 
Gaussian shape of $C({\bf q},{\bf K})$ since (\ref{1.6}) determines 
the curvature components of the correlator at ${\bf q} = 0$ \cite{CSH95}:
 \begin{equation}
   R_{ij}^2({\bf K}) = - \left. { \partial^2 C({\bf q},{\bf K})\over 
   \partial q_i\, \partial q_j}
        \right\vert_{ {\bf q} = 0}\, ,
 \label{1.6}
 \end{equation}
and these coincide with the experimentally determined half widths of 
$C({\bf q},{\bf K})$ only for Gaussian correlators. Non-Gaussian
characteristics of the correlator are difficult to quantify and
control in this Gaussian framework.

Here, we suggest to characterize the correlator with a different
set of a few HBT parameters, so-called $q$-moments, whose extraction 
does not presuppose a particular shape of $C({\bf q},{\bf K})$. 


The HBT parameters used in the Gaussian framework are based on 
space-time variances $\langle \tilde{x}_{\mu}\tilde{x}_{\nu}\rangle$, 
i.e. on expectation values $\langle f(x)\rangle$ of the emission 
function in coordinate space, cf.\ (\ref{1.3c}). In contrast, we 
suggest here to base a quantitative analysis of $C({\bf q},{\bf K})$ 
on $q$-variances which are expectation values $\langle\!\langle 
g({\bf q})\rangle\!\rangle$ of the true correlator $C({\bf q},{\bf K}) 
- 1$ in relative momentum space. 

For a Gaussian correlator, the HBT parameters can be determined either 
by fitting to the Gaussian (\ref{1.2}), or by computing the integrals
   \begin{eqnarray}
     {\langle\!\langle q_i\, q_j \rangle\!\rangle} &=& 
     {\int d^3q\,\, q_i\, q_j\,\, 
       {\lbrack{ C({\bf q},{\bf K}) - 1}\rbrack}
       \over \int d^3q\, {\lbrack{ C({\bf q},{\bf K}) - 1}\rbrack}} 
     = {\textstyle{1\over 2}}\, {\left({ R^{-1}({\bf K})}\right)}_{ij},
     \label{2.1} \\
        \lambda &=& \sqrt{\det R({\bf K})/\pi^3}
        \int d^3q\,
        {\lbrack{ C({\bf q},{\bf K}) - 1}\rbrack}\, .
     \label{2.2}
   \end{eqnarray}
For a non-Gaussian correlator, we can {\it define} the HBT radius
parameters and the intercept $\lambda$ in terms of these ``$q$-variances''.

The deviations of the correlator from a Gaussian shape (``non-Gaussicities'')
are then quantified by higher order $q$-moments. Due to the 
${\bf q}{\to}-{\bf q}$ symmetry of $C({\bf q},{\bf K})$ the first 
non-Gaussian contribution shows up in the fourth order $q$-moments 
(``kurtosis''). All higher order $q$-moments can be calculated as 
derivatives of the generating function $Z({\bf y},{\bf K})$,
  \begin{eqnarray}
    \label{2.3} 
    Z({\bf y},{\bf K}) &=& 
        \int d^3q\, e^{i{\bf q}\cdot {\bf y} }\,
                   {\lbrack{ C({\bf q},{\bf K}) - 1 }\rbrack}\, ,\\
    \label{2.4}
     {\langle\!\langle{q_{i_1}\,q_{i_2} ... q_{i_n}}\rangle\!\rangle} &=& 
     (-i)^n {\partial^n\over \partial\!y_{i_1} \partial\!y_{i_2} ...
     \partial\!y_{i_n}} \ln Z({\bf y,K})\bigg\vert_{{\bf y}=0} \, .
  \end{eqnarray}
>From this generating function, the correlator can be reconstructed 
completely. The series of $n$-th $q$-variances (\ref{2.4}) is merely
a convenient way to characterize its shape starting with its 
``Gaussian'' widths $\langle\!\langle q_i q_j \rangle\!\rangle$
and going for increasing $n$ step by step to finer structures. 

To calculate $Z({\bf y},{\bf K})$ directly from a given emission 
function it is convenient to use the normalized ``relative 
distance distribution''
 \begin{eqnarray}
   \rho(u;K) &=& \int d^4X\, s(X+{\textstyle{u\over 2}},K) \,
                           s(X-{\textstyle{u\over 2}},K) \, ,
   \nonumber\\
 \label{2.5}
   && \int d^4u\, \rho(u;K) = 1\, ,
 \end{eqnarray}
written in terms of the normalized emission function
$s(x,K) = S(x,K) / \int d^4x\, S(x,K)$. $\rho$ is real and even in $u$.
Then 
 \begin{equation}
 \label{2.6}
   Z({\bf y,K}) = \int d^3q \, e^{i{\bf q}{\cdot}{\bf y}}\, 
                  \int d^4u \, e^{iq{\cdot}u}\, \rho(u;K) \, .
 \end{equation}
In the Cartesian parametrization, this expression simplifies to a 
one-dimensional integral:
 \begin{equation}
 \label{2.7}
   Z({\bf y,K}) =
   \int dt\, \rho(y_s,y_o+\beta_\perp t, y_l+\beta_l t, t; K)\, .
 \end{equation}
For Gaussian emission functions Eqs.~(\ref{2.1}), (\ref{2.4}) and
(\ref{2.7}) reproduce the relations (\ref{1.5}). 


The method of $q$-variances generally requires to invert the matrix
(\ref{2.1}) and to discuss a 4-dimensional tensor of fourth order
moments. Such a full analysis will be published elsewhere~\cite{FKW97}.
Here, we restrict ourselves to a unidirectional analysis of the  
correlations $\tilde C(q_i,{\bf K}) \equiv C(q_i, q_{j\ne i}{=}0,{\bf K})$
along the three Cartesian axes. For the relative momentum component $q_i$, 
the corresponding HBT radius parameter and intercept are then defined via 
the relations (we use the same notation as for the three-dimensional
$q$-moments) 
  \begin{mathletters}
  \label{2.8}
  \begin{eqnarray}
  \label{2.8a}
     R_i^2({\bf K}) &=& {1\over 2\,\langle\!\langle q_i^2 
     \rangle\!\rangle}\, ,
  \\
  \label{2.8b}
     \langle\!\langle q_i^2 \rangle\!\rangle
     &=& {\int dq_i\, q_i^2\, [\tilde C(q_i,{\bf K})-1] \over
          \int dq_i\, [\tilde C(q_i,{\bf K})-1] }\, ,
  \\
  \label{2.8c}
     \lambda_i({\bf K}) &=& (R_i({\bf K})/\sqrt{\pi})
     \int dq_i\, [\tilde C(q_i,{\bf K})-1] \, .
  \end{eqnarray}
  \end{mathletters}
Even Gaussian correlators have non-vanishing higher $q$-moments:
 \begin{equation}
 \label{2.9}
   \langle\!\langle q_i^{2m}\rangle\!\rangle^{\rm Gauss} =
        {(2m - 1)!!\over (2R_i^2)^m} \, .
 \end{equation}
Being expressible in terms of the second moments they do not contain
new information. The non-trivial higher order information is contained
in the (normalized) $q$-cumulants, in which these trivial contributions
are subtracted:
  \begin{equation}
    \label{2.10}
        \Delta_i^{(2m)} = {1\over (2m-1)!!}\,
        { \langle\!\langle q_i^{2m} \rangle\!\rangle
          \over
          \langle\!\langle q_i^2 \rangle\!\rangle^m} - 1\,.
  \end{equation}
The normalization removes the dependence on the Gaussian widths of 
$C({\bf q},{\bf K})$ and turns them into dimensionless measures
for deviations from a Gaussian shape.

To extract the moments ${\langle\!\langle{q_i^n}\rangle\!\rangle}$ 
from data one replaces (\ref{2.8b}) by a ratio of sums over bins in the 
$q_i$-direction. The higher the order $n$ of the $q$-moment, the 
more sensitive are the extracted values to statistical and systematic
uncertainties in the region of large $q_i$. First investigations with 
event samples generated by the VENUS event generator indicate that the 
current precision of the data in the Pb-beam experiments at the CERN 
SPS permits to determine the second and fourth order $q$-moments~\cite{FKW97}.
Accordingly, we restrict our discussion of non-Gaussian features to
the ``kurtosis''
  \begin{equation}
  \label{2.11}
    \Delta_i = { \langle\!\langle q_i^4 \rangle\!\rangle
                 \over
                3\, \langle\!\langle q_i^2 \rangle\!\rangle^2} - 1\, .
  \end{equation}
%


We have computed the unidirectional Gaussian parameters $R_i$ 
and the kurtosis $\Delta_i$ for a model pion emission function
including resonance decay channels $R$: 
  \begin{equation}
     S_{\pi}(x,p) = S_{\pi}^{\rm dir}(x,p) + \sum_{R} S_{R\to \pi}(x,p)\, .
  \label{3.1}
  \end{equation}
The contributions $S_{R\to\pi}$ are obtained from
the direct resonance emission function $S_R^{\rm dir}(X,P)$  
by propagating the resonances of widths
$\Gamma$, produced at $(X_\mu,P_\mu)$, along a classical path 
$x^\mu = X^\mu + {P^\mu\over M} \tau$ according to an exponential
decay law~\cite{G77,Marb}:
 \begin{eqnarray}
   S_{R\to\pi}(x;p) &=& 
        M\, \int_{s_-}^{s_+} ds\, g(s)
        \int{d^3 P \over E_{_P}}\, 
        {\delta}{\left({P\cdot p - M E^*}\right)} \nonumber \\
   &&\!\times\! \int d^4X \, 
        \int d\tau \, \Gamma e^{-\Gamma\tau} \,
        \delta^{(4)}\textstyle{\left( x - 
            \left( X + {P\over M} \tau \right) \right)} \nonumber \\
   && \!\times\, S_R^{\rm dir}(X,P)\, .
 \label{3.2}
 \end{eqnarray}
We consider isotropic decays in the resonance rest frame with
decay phase space $g(s)$, $E^*$ being the energy of the observed decay 
pion in this frame and $s$ the squared invariant 
mass of the $(n-1)$ unobserved decay products.

Our model assumes local thermalization at freeze-out and produces 
hadronic resonances by thermal excitation. For particle species $i$ 
with spin degeneracy $2J_i+1$, the emission function reads \cite{CNH95}
 \begin{eqnarray}
 \label{3.3}
   S^{\rm dir}_i(x,P) &=& \textstyle{2J_i + 1 \over (2\pi)^3}\,
   P{\cdot}n(x)\,
   \exp{\left(- {P \cdot u(x) - \mu_i \over T} \right)}\,  H(x)\, ,
   \nonumber \\
   H(x) &=& 
          \exp\left( - {r^2\over 2 R^2} 
                     - {\eta^2\over 2 (\Delta\eta)^2}
                     - {(\tau-\tau_0)^2 \over 2 (\Delta\tau)^2}
                 \right) \, . 
 \end{eqnarray}
The Boltzmann factor $\exp[-(P{\cdot}u(x) - \mu_i)/T]$ 
implements both the assumption of thermalization, with temperature $T$ 
and chemical potential $\mu_i$, and collective expansion with 
hydrodynamic flow $4$-velocity $u_{\mu}(x)$. Space-time is parametrized 
via longitudinal proper time $\tau = \sqrt{t^2-z^2}$, space-time 
rapidity $\eta= {1\over 2} \ln{[(t+z)/(t-z)]}$, transverse radius $r$ 
and azimuthal angle $\phi$. The Gaussian factors in $H(x)$ specify 
the spatial extension of the source as well as a Gaussian average 
around a mean freeze-out proper time $\tau_0$ with dispersion $\Delta\tau$.
For freeze-out along proper time hyperbolas $P{\cdot}n(x)= M_\perp 
\cosh(Y-\eta)$ \cite{CNH95} where $Y$ and $M_\perp$ are the rapidity 
and transverse mass associated with $P$.

\vskip -1.0cm
\begin{figure}[h]\epsfxsize=10cm 
\centerline{\epsfbox{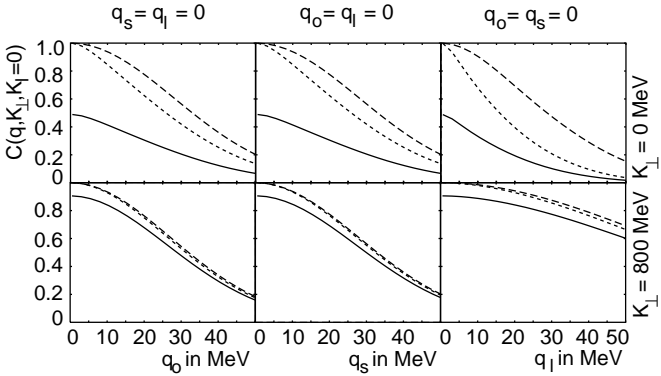}}
\vskip -6.5cm
\caption{\it 
Two-pion correlations for transverse flow $\eta_f = 0$. 
Curves are calculated for the model emission function 
(\protect\ref{3.3}) without
resonance decay contributions (dashed lines), including pions from 
the shortlived resonances $\rho$, $\Delta$, $K^*$, $\Sigma^*$ and
from the $\omega$ (dotted lines), and including also pions from the 
longlived resonances $\eta$, $\eta'$, $K_S^0$, $\Sigma$, $\Lambda$ 
(solid lines). 
}\label{fig1}
\end{figure}

In the longitudinal direction we assume scaling expansion, $v_l=z/t$
or $\eta_l = {1 \over 2} \ln[(1 + v_l)/(1-v_l)] = \eta$. For the
transverse expansion we choose a linear rapidity profile:
 \begin{equation}   
 \label{3.4}
   \eta_t(r) = \eta_f \left({r\over R}\right)\, .
 \end{equation}

All numerical results presented here are obtained for
the set of source parameters $T = 150$ MeV, $R = 5$ fm, $\Delta\eta = 1.2$,
$\tau_0 = 5$ fm/c, $\Delta\tau = 1$ fm/c and $\mu_B = \mu_S = 0$.
We include all pion decay channels of $\rho$, $\Delta$, $K^*$, 
$\Sigma^*$, $\omega$, $\eta$, $\eta'$, $K_S^0$, $\Sigma$ and $\Lambda$ 
with branching ratios larger than 5 percent~\cite{PDB94}.


For this model we have numerically computed the 2-pion correlator. 
Typical results are shown in Fig.~\ref{fig1}. A detailed model study 
of how resonance contributions affect the shape and pair momentum 
dependence of the correlator will be published elsewhere~\cite{WH96}.
Here, we merely observe that due to their exponential decay law, 
resonance decays can contribute exponential tails to the emission 
function $S_{R\to\pi}$ in (\ref{3.2}), thereby leading to a 
generically non-Gaussian shape of the correlator.
 
\begin{figure}[h]\epsfxsize=9cm 
\centerline{\epsfbox{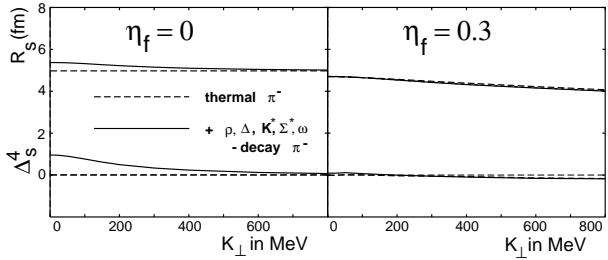}}
\vskip -6.5cm
\caption{\it The inverted $q$-variance $R_s$ of (2.8a) and the 
kurtosis $\Delta_s$ of (2.11) in the side direction at $Y=0$ as 
functions of $K_\perp$. Left: $\eta_f = 0$ (no transverse flow).
Right: $\eta_f = 0.3$. The difference between the dashed and solid 
curves is entirely dominated by $\omega$ decays.
}\label{fig2}
\end{figure}

To illustrate the use of $q$-variances we now discuss how the 
space-time characteristics of the emission region, specified in our 
model by the geometric input parameters and the transverse flow 
$\eta_f$, show up in these new 2-particle observables. First, we 
checked that for the model (\ref{3.1}-\ref{3.3}) the inverted 
second $q$-moments $R_i$ coincide in all three directions $i = 0,s,l$ 
with the Gaussian fit parameters extracted for a set of $n$ equidistant 
points $q_i^{(j)}$ between 0 and 50 MeV by minimizing  
  \begin{equation}
    \label{3.5}
    \sum_{j=1}^n {\left({ \ln \tilde C(q_i^{(j)},{\bf K})
                          -\log\lambda + R_i^2\, {q_i^{(j)}}^2
                          }\right)} = min. 
  \end{equation}
These ``Gaussian'' radius parameters $R_i$ and their transverse 
momentum dependence is well-studied for model emission functions 
(\ref{3.3}) not including resonance decay contributions~\cite{WSH96}. 
Especially, it is known that the transverse homogeneity length $R_t$ 
of (\ref{3.3}), i.e. the size of the effective emission region in the 
transverse plane, shrinks with transverse flow $\eta_f$ 
approximately like~\cite{WSH96}
   \begin{equation}
     \label{3.6}
     R_t \approx R {\left({ 1 + \textstyle{M_\perp\over T}\eta_f^2 
                            }\right)}^{-{1\over 2}}\, .
   \end{equation}
This introduces an $\eta_f$-dependent $K_\perp$-slope of $R_s$ which
was interpreted previously as a signature of transverse flow.
Fig.~\ref{fig2} shows that if resonance decay contributions are added
in the calculation, this distinctive behaviour is lost: $R_s$ develops a 
$K_\perp$-slope even for scenarios without transverse flow.
The reason is that resonances can propagate outside of the thermally
equilibrated region before decaying. This effect (``lifetime effect'') 
leads to exponential tails in $S_{R\to\pi}$ and tends to increase the 
Gaussian widths $R_i$. Since the relative abundance of resonances is 
larger for small $K_\perp$, this tail is more pronounced in the region
of small $K_\perp$, thereby leading to a $K_\perp$-slope of $R_s$.
 
In contrast, for a finite transverse flow $\eta_f = 0.3$ the 
$K_\perp$-dependence of $R_s$ is due to the flow, and resonance decay 
contributions do not change the slope of $R_s$ in our model. This can be
traced back to the shrinking transverse size (\ref{3.6}) of the direct 
resonance emission function $S_R^{dir}$ in (\ref{3.3}): due to their 
larger rest mass, parent resonances have a smaller effective emission 
region than thermal pions in the transverse direction. For the case 
depicted in Fig.~\ref{fig2}, this effect counterbalances the lifetime 
effect almost exactly.

According to (\ref{3.6}), the absolute size of the {\it side} radius 
depends not only on $\eta_f$ but also on the input parameter $R$ for
which no independent measurement exists. As a consequence, the Gaussian
radius parameter $R_s(K_\perp)$ by itself does not allow to distinguish
scenarios with and without transverse flow once resonance decays are 
included. This conclusion extends to the other Gaussian widths, $R_o$ and 
$R_l$, as well.

\vskip -1.5cm

However, the physical origin of the $K_\perp$-slope of $R_s$ is different 
for the two situations shown in Fig.~\ref{fig2}. This shows up in the 
kurtosis of the corresponding correlators: without transverse flow,
resonance decay contributions increase $R_s$ due to the lifetime effect 
which generates exponential tails in $S_{R\to\pi}$. These tails manifest 
themselves in the correlator through a significant non-Gaussian component. 
For non-zero transverse flow, on the other hand, the $K_\perp$-slope 
arises from the $K_\perp$-dependent shrinking (\ref{3.6}) of the 
effective transverse emission region. This effect is more prominent for
resonances than for thermal pions, i.e. $S_\pi^{dir}$ is spatially more 
extended in the transverse plane than $S_R^{dir}$, and hence it ``covers'' 
a substantial part of the exponential tails of $S_{R\to\pi}$. As a 
consequence, the total emission function (\ref{3.1}) can be expected 
to show much smaller deviations from a Gaussian shape for the scenario 
with transverse flow, and this again leads to a more Gaussian correlator.

This explains why the kurtosis $\Delta_s(K_\perp)$ plotted in 
Fig.~\ref{fig2} provides a clearcut distinction between the two 
scenarios. $\Delta_s$ simply reflects the importance of resonance 
decays. For vanishing transverse flow, $\Delta_s$ is significant 
and decreases with increasing $K_\perp$ since the relative
abundance of resonance decay pions dies out; for finite transverse flow
$\eta_f = 0.3$, it is an order of magnitude smaller. A $K_\perp$-dependence
of $R_s$ without measurable kurtosis is thus a clear sign of transverse
flow.

Here, we do not discuss to what extent this particular $\eta_f$-dependence 
of the kurtosis $\Delta_s$ is generic for realistic models of heavy ion 
collisions including resonance decays. This will require a more systematic
study \cite{WH96}. We simply conclude that higher order $q$-variances 
can provide crucial additional information on the correlator 
which can help to distinguish physical scenarios which are difficult 
to disentangle on the level of ``Gaussian'' HBT radius parameters. On the
other hand, they can still be directly related to specific features of the 
emission function and thus should aid the spatio-temporal interpretation
of correlation measurements. This is less straightforward if one
only knows the overall shape of the correlator, without decomposing it into
its moments.

In summary, we expect these new HBT parameters, both the one- and 
multi-dimensional $q$-moments of the correlator, to become valuable 
additional observables in future comparisons between models and experiment.

This work was supported by BMBF, DFG and GSI. Discussions with
P. Foka, H. Kalechofsky, B. Lasiuk, M. Martin, H.-P. Naef, L. Rosselet,
P. Seyboth and S. Voloshin are gratefully acknowledged.


\end{document}